%% file: efficient-lightweight-client-blockchain.tex
\documentclass{IEEEtran}

\usepackage{breakurl}
\usepackage{doi}
\usepackage{amsmath}
\usepackage{amssymb}
\usepackage{graphicx}
\usepackage{url}
\usepackage{color}
\usepackage{float}
\usepackage{algorithm,algorithmic}
\usepackage{balance}

\newtheorem{thm}{Theorem}

\begin{document}

\title{Efficient Public Blockchain Client for Lightweight Users
\thanks{A preliminary version of this paper was published in SERIAL'17}
}

\author{Lei Xu, Lin Chen, Zhimin Gao, Shouhuai Xu, Weidong Shi\\
xuleimath@gmail.com, chenlin198662@gmail.com, mtion@msn.com, shouhuai.xu@utsa.edu, wshi3@central.uh.edu}

\maketitle

\begin{abstract}
Public blockchains provide a decentralized method for storing transaction data and have many applications in different sectors. In order for users to track transactions, a simple method is to let  them keep a local copy of the entire public ledger. Since the size of the ledger keeps growing, this method becomes increasingly less practical, especially for lightweight users such as IoT devices and smartphones. In order to cope with the problem, several solutions have been proposed to reduce the storage burden. However, existing solutions either achieve a limited storage reduction (e.g., simple payment verification), or rely on some strong security assumption (e.g., the use of trusted server). In this paper, we propose a new approach to solving the problem. Specifically, we propose an \underline{e}fficient verification protocol for \underline{p}ublic \underline{b}lock\underline{c}hains, or EPBC for short.
EPBC is particularly suitable for lightweight users, who only need to store a small amount of data that is {\it independent of} the size of the blockchain. We analyze EPBC's performance and security, and discuss its integration with existing public ledger systems. Experimental results confirm that EPBC is practical for lightweight users.
\end{abstract}

\section{Introduction}
A public blockchain or ledger consists of a set of blocks that are linked together, where each block contains a set of transactions. A public blockchain is maintained by a group of users, who run a consensus protocol (e.g., proof-of-work with longest-chain) to resolve disagreements regarding the blockchain. In a simple realization of public blockchain, each user keeps a {\em local} copy of the entire blockchain, meaning that each user has access to all historic activities and can easily test whether a new transaction is consistent with the existing transactions. This explains why a public ledger does not have to rely on any centralized party. This technique is central to many popular applications, such as Bitcoin \cite{nakamoto2008bitcoin}.

Although keeping a local copy of the blockchain in question simplifies many operations (e.g., transaction searching and balance calculation), this imposes a substantial storage overhead because the blockchain keeps growing. For example, the Bitcoin blockchain includes 472,483 blocks in June 2017, or 120 GB in volume. This overhead may not be a problem for modern servers and PCs, but are prohibitive for lightweight users such as mobile devices and IoT devices. In general, this would hinder the development of applications that aim are meant to be built on top of blockchains (e.g., smart contract system~\cite{wood2014ethereum}). At the same time, smart phones are the major way to get online in some areas, especially in underdeveloped countries, and there is a big need for mobile and lightweight users to use blockchains~\cite{christidis2016blockchains}.
Therefore, it is urgent to reduce the storage overhead, especially for those lightweight users.

Indeed, Nakamoto proposes the simplified payment verification (SPV) protocol in the very first Bitcoin paper~\cite{nakamoto2008bitcoin}, which requires a client to store {\em some}, instead of all, blocks while being able to check the validity of transactions recorded in the blockchain. This technique is also widely used in many blockchain-based applications, such as smart contract system~\cite{wood2014ethereum}. The basic idea underlying the SPV protocol is that each user only needs to keep the headers of blocks, rather than the blocks themselves. This means that the local storage overhead still increases linearly with the number of blocks, which grows over time and can quickly become prohibitive for lightweight users. An alternate approach is that a lightweight user chooses to trust some nodes in a blockchain system. However, this practice sacrifice the most appealing feature of the blockchains, namely the absence of any trusted third party. Moreover, this approach can be vulnerable to, for example, Sybil attacks~\cite{douceur2002sybil}.

In this paper, we propose an {\em efficient  verification protocol for public blockchain}, dubbed EPBC. The core of EPBC is a succinct blockchain verification protocol that ``compresses'' the whole chain to a constant-size summary, using a cryptography accumulator~\cite{benaloh1993one}. A lightweight user only needs to store the most recent summary, which is sufficient for the user to verify the validity of transactions. EPBC can be incorporated into existing blockchains as a middle layer service, or can be seamlessly incorporated into new blockchain systems.

In summary, our contributions in this work include:
\begin{itemize}
\item We design a novel scheme for lightweight users to use public blockchains using cryptographic accumulator.
\item We analyze the security and asymptotic performance of the scheme, including its storage cost.
\item We report a prototype implementation of the core protocol of EPBC and measure its performance. Experimental results show that the scheme is practical for lightweight users.
\end{itemize}

The rest of the paper is organized as follows.
In Section~\ref{sec-background} we briefly review the background of public blockchains and the simplified payment verification protocol.
In Section~\ref{sec-core-component} we describe the design of the core component of EPBC, i.e., efficient block verification, and analyze its security.
Section~\ref{sec-convert-operations} describes two common operations for blockchain based applications using the core component of EPBC, and we provide the architecture to integrate EPBC with existing blockchain systems in Section~\ref{sec-integration}.
Experimental results are given in Section~\ref{sec-imp-exp} to demonstrate the practicability of EPBC, and Section~\ref{sec-related} discusses the related prior work.
We conclude the paper in Section~\ref{sec-conclusion}.

\section{Background of Public Blockchain}\label{sec-background}
A blockchain is a distributed ledger that has been used by Bitcoin and other applications to store their transaction data, where a transaction can be a payment operation, smart contract submission, or smart contract execution result submission. There are different approaches to construct blockchains. In this work, we focus on the class of blockchains that are built on the principle of proof-of-work (PoW)~\cite{vukolic2015quest}. This class of blockchains have a low throughput and a high latency, but have the desirable properties of fairness and expensive-to-attack. Furthermore, there are many efforts at improving their performance~\cite{croman2016scaling,luu2015scp} and characterizing their security properties~\cite{gervais2016security}.

\begin{figure}[!htbp]
  \centering
  \includegraphics[width=3.5in]{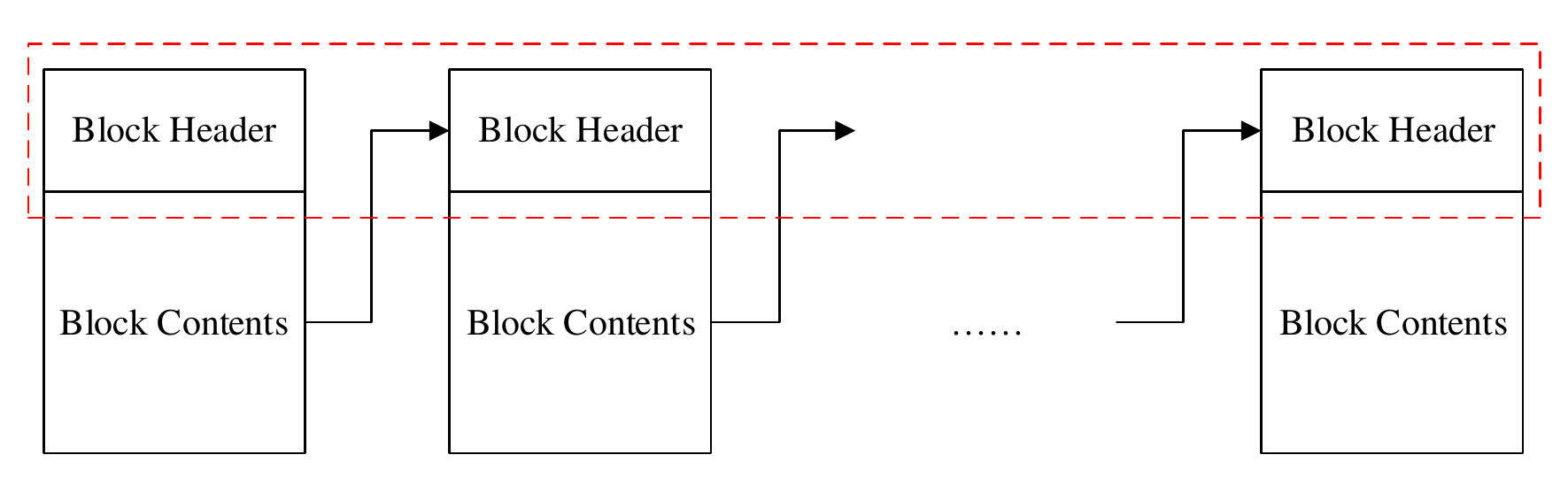}
  \caption{In the SVP scheme, a user stores the headers of the blocks, rather than the blocks themselves. A header contains the relevant meta data (e.g., the root of the Merkle tree whose leaves are the transactions contained in a block). This allows a user to verify whether a given block is valid or not.}
  \label{fig-svp}
\end{figure}

Since a blockchain is immutable and append-only, its size keeps growing. There are proposals for coping with this issue. A straightforward approach is to trust some user, who can check the validity of transactions on the user's behalf. This approach assumes that the lightweight user always knows who can be trusted. Another approach is to use the SPV protocol mentioned above~\cite{nakamoto2008bitcoin}. In this scheme, as highlighted in \figurename~\ref{fig-svp}, a user only needs to store the block headers, which contain the root of the Merkle tree of the transactions in the corresponding block. When a user needs to verify a transaction, it sends a request to the system asking for the corresponding block, whose validity can be verified by using the root of the Merkle tree.

\section{Design and Analysis of EPBC}\label{sec-core-component}

\subsection{Design Objective and Assumption}
The objective of EPBC is to allow lightweight users to participate in applications that use public blockchains.
By ``lightweight users'' we mean the users who use devices that have limited computation/storage capacities, such as IoT devices and smartphones.
Specifically, EPBC aims to allow lightweight users to achieve the following:
\begin{itemize}
\item {\em Efficient storage}: A user does not have to store or download the entire blockchain. Instead, a user only needs to consume a storage that is ideally independent of the size of the blockchain.
\item {\em Verifiability of transactions}: A user can verify whether a transaction has been accepted by the blockchain or not.
\end{itemize}

Like any public blockchain constructed according to proof-of-work, we assume that the majority of the users are honest.

In what follows, we first describe the block verification protocol, which is the core component of EPBC. Then, we describe how to use this protocol to construct the EPBC scheme.

\subsection{The Block Verification Protocol}

\begin{figure}
  \centering
  \includegraphics[width=3in]{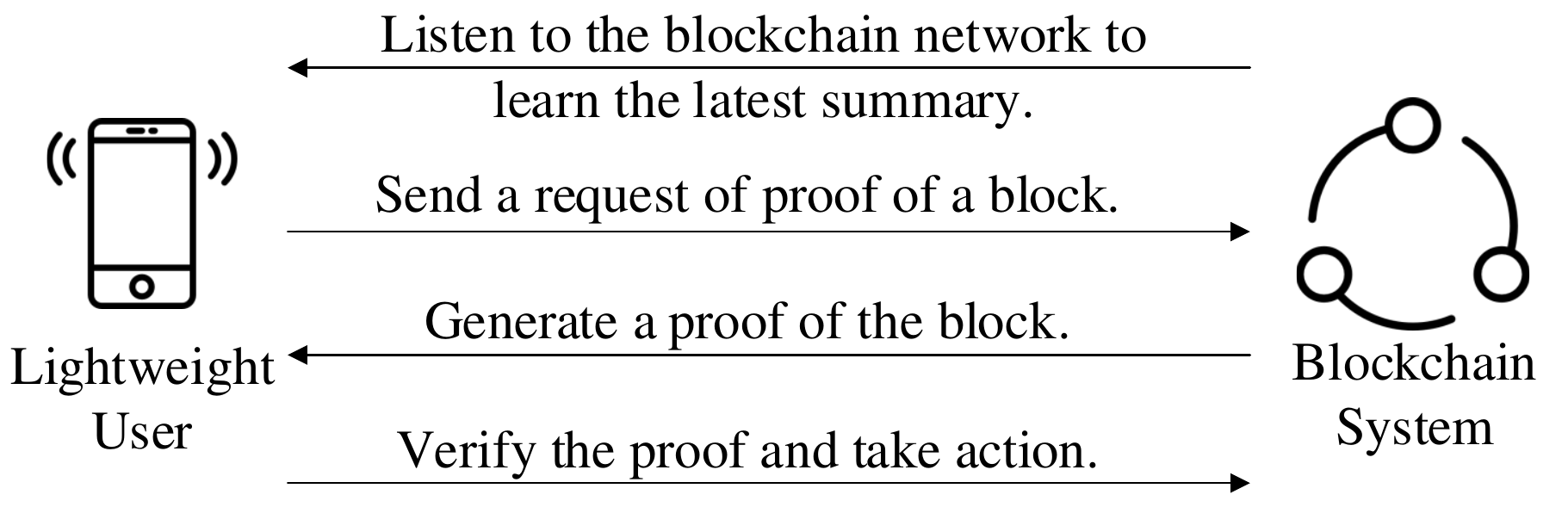}
  \caption{Illustration of the blockchain verification protocol. The nodes in the blockchain system with bigger storage capacities can keep a full copy of the blockchain. These nodes will interact with the lightweight users to help the latter to verify the validity of blocks.}
  \label{fig-overview}
\end{figure}

\figurename~\ref{fig-overview} gives an overview of the verification protocol.
Basically, a lightweight user can verify the validity of transactions by interacting with the blockchain system.

The blockchain verification protocol of EPBC consists of the following four algorithms:
\begin{itemize}
\item {\it Setup}: This algorithm is executed once by the creator of the blockchain. The algorithm generates the public parameters that are needed by the other algorithms.
\item {\it Block and summary construction}: This algorithm generates blocks and a summary of the current blockchain. Anyone participating in the mining competition to build new blocks is responsible for calculating the summary of the current blockchain. The summary depends on the content of the current blockchain and the public parameters.
\item {\it Proof generation}: This algorithm generates a proof for a given block. The proof may depend on, among other things, the entire blockchain.
\item {\it Proof verification}: Given the summary of a blockchain and a proof for a single block, this algorithm verifies whether the proof is valid or not.
\end{itemize}
With this protocol, a lightweight user keeps the updated summary of the blockchain. When the user wants to verify a specific block, it can ask the parties that are involved in a transaction for a proof for the block, which is generated by running the {\it proof generation} algorithm. The user then executes the {\it proof verification} algorithm to determine whether to accept the block or not. In what follows we describe the details of these algorithms.

\paragraph{Setup.}
The creator of the blockchain selects two large prime numbers $p,q$, and calculates $N=pq$ as in the RSA accumulator system. $N$ is embedded into the first block and disclosed to the public; and then $p,q$ are discarded. The creator also selects a random value $g \in \mathbb{Z}_N^*$. Each block will be labelled with an integer, with the ``genesis'' block (i.e., the first block on the blockchain) has the label ``1''.

\paragraph{Block and summary construction.}
Each block contains, in addition to the standard attributes (e.g., transaction information and proof-of-work nonce), a new attribute $S$, which is the summary of the current blockchain. For the $i$-th block, which is denoted by $\mathit{blk}_i$, the attribute $S_i$ is calculated and stored with $\mathit{blk}_i$ as follows:
$$
S_i =
\begin{cases}
g^{\texttt{hash}(\textit{blk}_i || i)} \mod N, &\text{if $i=1$},\\
S_{i-1}^{\texttt{hash}(\textit{blk}_i || i)} \mod N, &\text{if $i>1$}.\\
\end{cases}
$$
If the current blockchain contains $n$ blocks, $S_n$ is the summary of the current blockchain. The block position information $i$ is used in the computation for the purpose of preventing the attacker from manipulating the position of a block. After the newly generated block is broadcast to the blockchain system, the following two algorithms can be executed.

\paragraph{Proof generation.}
To generate a proof that shows block $\textit{blk}_i$ is the $i$-th block on the blockchain with summary $S_n$, where $i \leq n$, the prover calculates $p_i = (p_i^{(1)},p_i^{(2)})$ as follows:
$$p_i =
\begin{cases}
  p_i^{(1)} &= \texttt{hash}(\textit{blk}_i || i),\\
  p_i^{(2)} &= g^{ (\prod_{k=1}^{n} \texttt{hash}(\textit{blk}_k || k)) / \texttt{hash}(\textit{blk}_i || i)} \mod N.
\end{cases}
$$
Note that the proof is generated by a user who keeps the entire blockchain and therefore can compute $p_i^{(2)}$ without knowing $\phi(N)$, where $\phi$ is the Euler function.

\paragraph{Proof verification.}
Given a block $\textit{blk}$, a claimed proof $p=(p^{(1)}, p^{(2)})$, and a blockchain summary $S_n$,
a user can verify that block $\textit{blk}_i$ is indeed the $i$-th block on blockchain with summary $S_n$, where $i \leq n$, as follows:
$$
\begin{cases}
  p^{(1)}  \stackrel{?}{=}  \texttt{hash}(\mathit{blk}_i || i),\\
  S_n      \stackrel{?}{=}  (p^{(2)} )^{p^{(1)}} \mod N.
\end{cases}
$$
If both equations hold, the user accepts that $p$ is a valid proof for $\textit{blk}$;
otherwise, the verifier rejects the block.

\subsection{Parameter Initialization}
One of the key steps in the blockchain verification protocol is the parameter initialization, i.e., selecting $p$ and $q$ to generate the modulus $N$. If $p$ or $q$ is exposed, the protocol is clearly not secure.
This issue can be addressed by generating $N$ using a multi-party protocol. There have been many protocols for this purpose. For example, the protocol proposed by Cocks \cite{cocks1997split} works as follows. Suppose at the beginning there are $\ell$ users who work together to generate the first block.
\begin{enumerate}
  \item Each user $i$, $1\leq i \leq \ell$, selects his/her own prime numbers $p_i,q_i$.
  \item Each user $i$, $1\leq i \leq \ell$, calculates $N=(p_1+p_2+\cdots + p_\ell) (q_1+q_2+\cdots + q_\ell)$. By leveraging the protocol given in~\cite{cocks1997split}, user $i$ can calculate $N$ without knowing the two factors of $N$.
  \item Each user tests whether $N$ is a product of two prime numbers or not. Specifically, the system randomly selects a random number $x$ and each user calculates $x^{p_i+q_i} \mod N$. If $\prod x^{p_i+q_i} \mod N \equiv x^{N+1} \mod N$, $N$ passes the test. Carmichael numbers that can pass this test can be further eliminated by methods given in~\cite{boneh1997efficient}.
  \item If the current $N$ passes all tests, users work together to embedded it in the genesis block. Otherwise they repeat the process again, until an appropriate $N$ is found.
\end{enumerate}
Since $N$ only needs to be generated once, the cost of the parameter initialization is not a big concern.

\subsection{Security and Performance of the Block Verification Protocol}
It is straightforward to verify that the protocol is correct, meaning that any legitimate proof will be accepted as valid.
The following theorem shows that for a given summary $S$ of blockchain $\mathit{BC}$, no attacker can generate a valid proof for a forged block $\mathit{blk}'$ that is not contained in $\mathit{BC}$ under strong RSA assumption.

\begin{thm}
Given a summary $S_n$ of blockchain $\mathit{BC}$, there is no probabilistic polynomial-time attacker $\mathcal{A}$ that can forge a block $\mathit{blk}'$ and an accompanying proof $P'$ that $\mathit{blk}'$ is a valid block on blockchain $\mathit{BC}$ in the random oracle model; otherwise, the Strong RSA assumption is broken.
\end{thm}

\begin{IEEEproof}
Suppose $\texttt{hash}()$ behaves like a random oracle.
Let $r_i = \texttt{hash}(\mathit{blk}_i || i)$ where $\mathit{blk}_i$ is the $i$-th block on $\mathit{BC}$, and $S_n = g^{\prod_{k=1}^{n} r_k} \mod N$.
We consider two scenarios of attacks:
\begin{itemize}
\item The attacker knows the summary $S_n$ but not the blockchain. Suppose the attacker chooses $\mathit{blk}'$ and position $i'$ for the block. Then, the attacker needs to compute $y \in \mathbb{Z}_n^*$ such that
        $$y ^ {\texttt{hash}(\mathit{blk}' || i')} \mod N = S_n.$$
    This immediately breaks the Strong RSA assumption.
\item The attacker knows both blockchain and the summary $S_n$. In this case, the attacker knows all valid proofs for blocks in $\mathit{BC}$, i.e., $(r_i, S_n^{\frac{1}{r_i}} \mod N),~i=1, \ldots, n$. Suppose the attacker can generate a valid proof for a forged block $\mathit{blk}'$ for some position $i'$. Let $r' = \texttt{hash}(\mathit{blk}' || i')$. If $r' | \prod_{i=1}^n r_i$, the attacker can successfully make a valid proof for $\mathit{blk}'$ at position $i'$ because the attacker can compute $(r', S_n^{\prod_{i=1}^n r_i / r'})$. Because the attacker cannot control the output of \texttt{hash()}, the probability that the attacker can succeed is equivalent to the probability that a random number $r'$ is a factor of another random number $R=\prod_{i=1}^n r_i$.
    According to Erd\"{o}s-Kac theorem~\cite{erdos1940gaussian} and its extension counting multiplicities~\cite{billingsley1969central}, the number of prime factors of $R$ counting multiplicity is $\mathcal{O}(\log \log R)$. With Binomial theorem, the total number of divisors of $R$ is $\mathcal{O}(2^{\log \log R}) = \mathcal{O}(\log R)$, and $\lim_{R \to \infty} \frac{\log R}{R} =0$. Therefore, the probability that an attacker can find $r'$ is negligible when $R$ is large enough.
    As long as the attacker cannot find such $r'$, a successful attack implies that the the Strong RSA assumption is broken.
\end{itemize}
In summary, there is no practical attack against the protocol in the random oracle model unless the Strong RSA assumption is broken.
\end{IEEEproof}

Performance of the major algorithms is analyzed as follows.
\begin{itemize}
\item {Block construction.} When compared with the straightforward method by which each user keeps the entire blockchain, our method incurs some extra work in the block construction algorithm. The extra work consists of two parts: evaluating the hash value of the new block and calculating the new summary. The computation overhead is constant (i.e., one hash calculation and one modular exponentiation) and the storage overhead is also constant (i.e., an element in $\mathbb{Z}_N$ for the summary). The summary also incurs extra communication cost, which is however small (e.g., 2048 bits for a 2048-bit $N$).
\item {Proof generation.} The proof generation algorithm does not incur extra storage. The computational cost is proportional to the length of the current blockchain (i.e., the number of blocks in the chain) and the position of the block. Suppose the length of the blockchain is $n$, and the proof of $i$-th block needs to be generated, where $i \leq n$. The prover needs to conduct one hash evaluation of the $i$th block, and calculates the product of hash values of blocks  $1,\ldots,i-1,i+1,\ldots,n$. In summary, the prover calculates $n+1-i$ hashes, $n-1$ multiplications, and one modular exponentiation. Since the nodes with sufficient storage capacity (rather than the lightweight users) are supposed to generate proofs, the protocol is practical.
\item {Proof verification.} The computational cost to verify the proof of a block includes one hash evaluation and one modular exponentiation, which is constant. This explains why the protocol is suitable for lightweight users who only keep the summary of the blockchain.
\end{itemize}

\subsection{Reducing Cost of Proof Generation}\label{subsec-tree-structure}
Although both the cost of updating the summary of a blockchain and the cost of verifying a block are constant, the computational complexity for the prover to generate a proof is $\mathcal{O}(n)$, where $n$ is the number of {\em current} blocks on the blockchain (i.e., $n$ keeps increasing). In the worst-case scenario, the prover needs to traverse all of the blocks on the blockchain to calculate the second part of the proof, namely
$$g^{ (\prod_{k=1}^{n} \texttt{hash}(\textit{blk}_k)) / \texttt{hash}(\textit{blk}_i)} \mod N.$$
In order to reduce the computational complexity incurred by this, we design a scheme that improves the computational efficiency at the price of a slight increase in storage.

\begin{figure}[!htbp]
  \centering
  \includegraphics[width=3.5in]{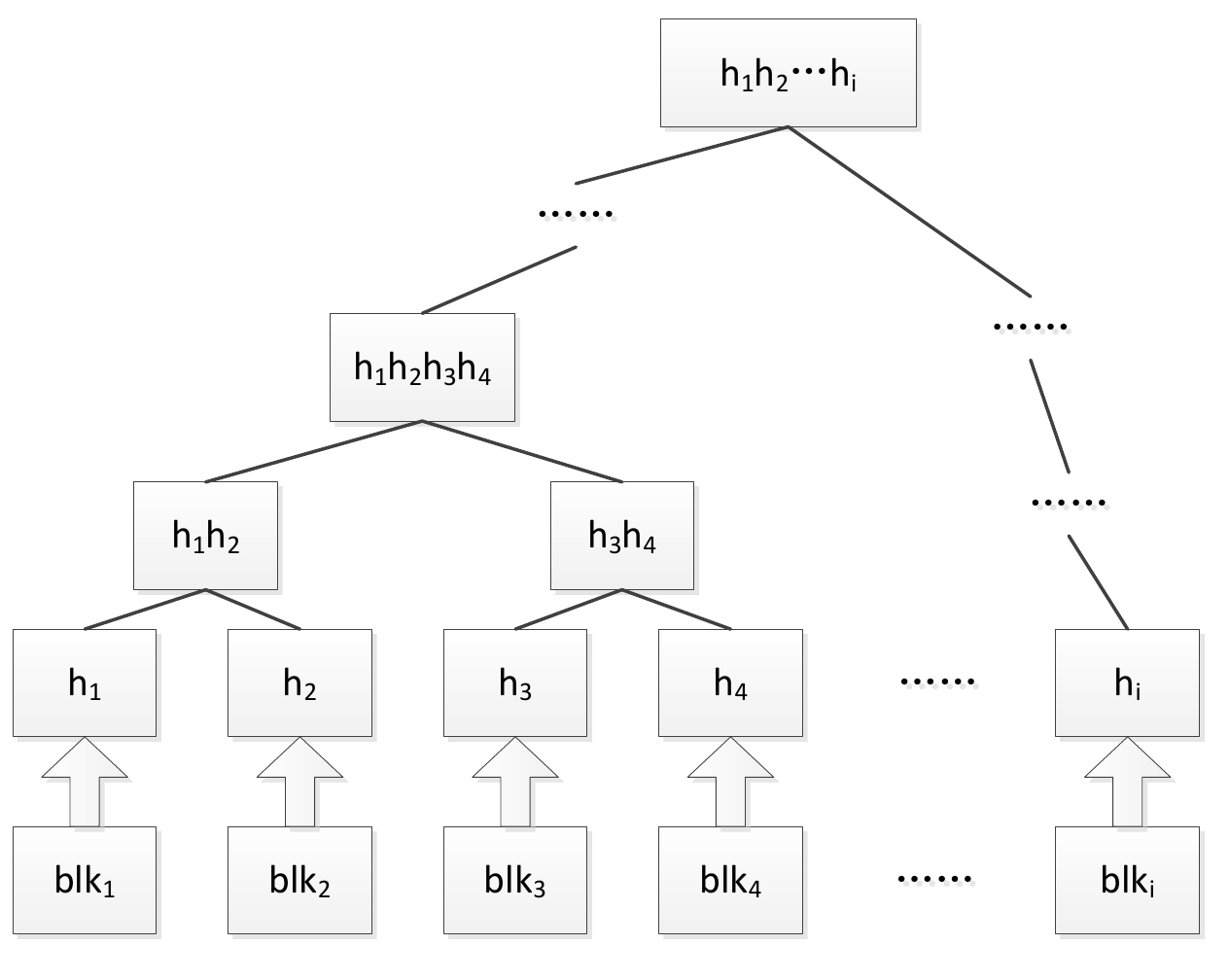}
  \caption{The storage structure that can be used by a prover to reduce its computational complexity when generating proofs. Each leaf $h_i$ stores the hash value of a block, and each internal node stores the product of the values stored at its two children.}
  \label{fig-accelerate-tree}
\end{figure}

\paragraph{Proof generation with a smaller computational complexty.}
The basic idea underlying the scheme is to let the prover maintain a binary tree $\mathcal{T}$. As illustrated in \figurename~\ref{fig-accelerate-tree}, the binary tree is used to store intermediate results that can be used to generate a proof for a given block. Specifically, each leaf stores the hash value of a corresponding block, and each internal node stores the product of its two direct children nodes. This way, the root node stores the product of the hash values of all of the blocks on the blockchain. The height of $\mathcal{T}$ is pre-determined. If a leaf is empty (i.e., currently there is no corresponding block on the blockchain), its value is set to 1 so that it does not contribute to the value stored at the root node.

Suppose the height of tree $\mathcal{T}$ is $h$ and the number of currrent blocks on blockchain is $n$, where $n < 2^{h-1}$. To calculate a proof for block $\mathit{blk}_i$, where $1 \leq i \leq n$, the prover leverages the information stored in $\mathcal{T}$ as follows:
\begin{itemize}
  \item Find the product of all of the values on the right-hand of $\mathit{blk}_i$ (the blockchain grows from left to right)
        \begin{equation}\label{equ-right}
          r \leftarrow \prod_{k=i+1}^{n} \texttt{hash}(\mathit{blk}_k).
        \end{equation}
        Instead of conducting the multiplication operation one-by-one, the prover utilizes different products information stored in $\mathcal{T}$ to accelerate the computation.
  \item Calculate $\mathit{LR} \leftarrow (S_i)^{r / \texttt{hash}(\mathit{blk}_i)} \mod N$.
  \item Set the proof as $P \leftarrow (\texttt{hash}(\mathit{blk}_i), LR)$.
\end{itemize}

Note that the height of $\mathcal{T}$ determines the number of blocks it can accommodate, and is therefore a pre-determined public parameter. If the height of $\mathcal{T}$ is $h$, the total number of blocks it can accommodate is $2^{h-1}$. This is no significant constraint because a relatively small $h$ can accommodate a large number of blocks. For example, when $h=32$, the structure can accommodate 4,294,967,296 blocks, which are about 9,000 times larger than the number of blocks on the Bitcoin network as of April 2017.

\paragraph{Analysis of the improved scheme.}
The improved scheme involves a binary tree $\mathcal{T}$ to store some information that can be used for generating proofs. Let $\texttt{height}(\mathcal{T})=h$, meaning that $n = 2^{h-1}$ is the number of leaves. Let $|\texttt{hash}()| = \ell$. At the leaf level (i.e., the first level), the size of each node is $\ell$. Each node at $i$-th level incurs $i \cdot \ell $ bits of storage, and the size of the root node is $h \cdot \ell$ bits. Therefore, the size of $\mathcal{T}$ is
\begin{align*}
& \underbrace{n \cdot \ell \vphantom{(n/2^i) \cdot (2^i \ell)} }_{\text{first level}}
\hphantom{~}+ \cdots +\hphantom{~}
\underbrace{(n/2^i) \cdot (2^i \ell)}_{\text{$i$-th level}}
\hphantom{~}+ \cdots +\hphantom{~}
\underbrace{ (n/2^{h-1}) \cdot (2^{h-1} \ell)}_{\text{$h$-th level, root}} \\
=&  \sum_{i=0}^{h-1} n \cdot \ell = h \cdot n \cdot \ell = (\log_2 n + 1) \cdot n \cdot \ell = \mathcal{O}(n\log n).
\end{align*}
With intermediate results stored in $\mathcal{T}$, the computation complexity for generating a proof is reduced to $h$ (or $\mathcal{O}(\log n)$) modular exponentiations.

More generally, if each internal node in \figurename~\ref{fig-accelerate-tree} has $m$ children, the height of $\mathcal{T}$ is reduced to $\log_m n + 1$. A similar analysis shows that the total size of $\mathcal{T}$ is $(\log_m n + 1) \cdot n \cdot \ell$, which is the size of storage a prover keeps locally. In order to calculate $r$, which is defined in Equation~(\ref{equ-right}), it requires about $\log_m n + m$ multiplication operations in the worst-case scenario, where $m$ is the number of multiplications incurred at an internal node at the second level of $\mathcal{T}$.
In order to select the value of $m$ so as to minimize the overall computational complexity, we calculate the derivative as follows:
\begin{equation*}
(\log_m n + m)' = (\frac{\ln n}{\ln m} + m)' = 1- \frac{\ln n}{m \ln^2m},
\end{equation*}
which monotonically increases with respect to $m$. Therefore, we get the minimum value when
$$
1 = \frac{\ln n}{m \ln^2m},
$$
and $m \approx \ln n$. In practice, we can set the number of branches to a small constant integer so as to reduce the computational complexity of the prover.

\section{Using the Block Verification Protocol to Construct EPBC}\label{sec-convert-operations}
In this section, we discuss construction of high-layer operations based on the verification protocol described in Section~\ref{sec-core-component}.
Specifically, we focus on two basic protocols: {\em blockchain identification} and {\em transaction verification}.

\paragraph{Blockchain identification.}
When a lightweight user needs to join a blockchain based application, it needs to obtain the current summary of the blockchain. Protocol~\ref{protocol-blockchain-iden} is for this purpose.

\begin{algorithm}\floatname{algorithm}{Protocol}
  \caption{Blockchain identification.}
  \label{protocol-blockchain-iden}
  \begin{algorithmic}[1]
    \STATE The lightweight user randomly selects a group of $\ell$ users, denoted by $G_u$, from the blockchain network;
    \FORALL{$u \in G_u$}
      \STATE The lightweight user queries $u$ to get the summary value $S^{(u)}$;
      \STATE The lightweight user interacts with $u$ to verify the validity of $S^{(u)}$ with respect to a random set of blocks chosen by the lightweight user;
    \ENDFOR
    \STATE The lightweight user calculates $$S \leftarrow \texttt{SummaryDetermination}(S^{(1)}, \ldots S^{(\ell)}),$$
    which returns the summary that is provided by majority of the users, where $S$ is the final summary of the blockchain;
  \end{algorithmic}
\end{algorithm}
Note that as long as the attacker does not control majority of the $\ell$ users, the protocol is secure. The lightweight user can also adopt other strategies to determine the summary, e.g., giving different weights to selected users and include this information when making the decision.

\paragraph{Transaction verification.}
A transaction is valid if and only if the block it belongs to is accepted by the majority of users, i.e., on the longest branch of the blockchain. Therefore, verification of a transaction is reduced to checking the validity of a block and its position (i.e., block number). A lightweight user can use the block verification protocol to verify that the block in question indeed contains the transaction in question. Then, the lightweight user can check the number of blocks that have been added after the block that is verified. Similar to the Bitcoin system~\cite{nakamoto2008bitcoin}, if more than 6 blocks have been added to the blockchain after the block under consideration, the transaction in question can be accepted with high confidentiality.

If the transaction is a smart contract submission or one-time smart contract execution result submission, the above method is also sufficient.
However, if the transaction is a payment operation or submission of multiple-time smart contract execution result, freshness becomes a concern. For example, the attacker can provide proof of an old block that contains previous payment of the same value. To prevent such attacks, the lightweight user can maintain a local counter and include the counter in its transactions.

\section{Integration with Existing Blockchain Systems}\label{sec-integration}
\begin{figure}[!htbp]
  \centering
  \includegraphics[width=3.5in]{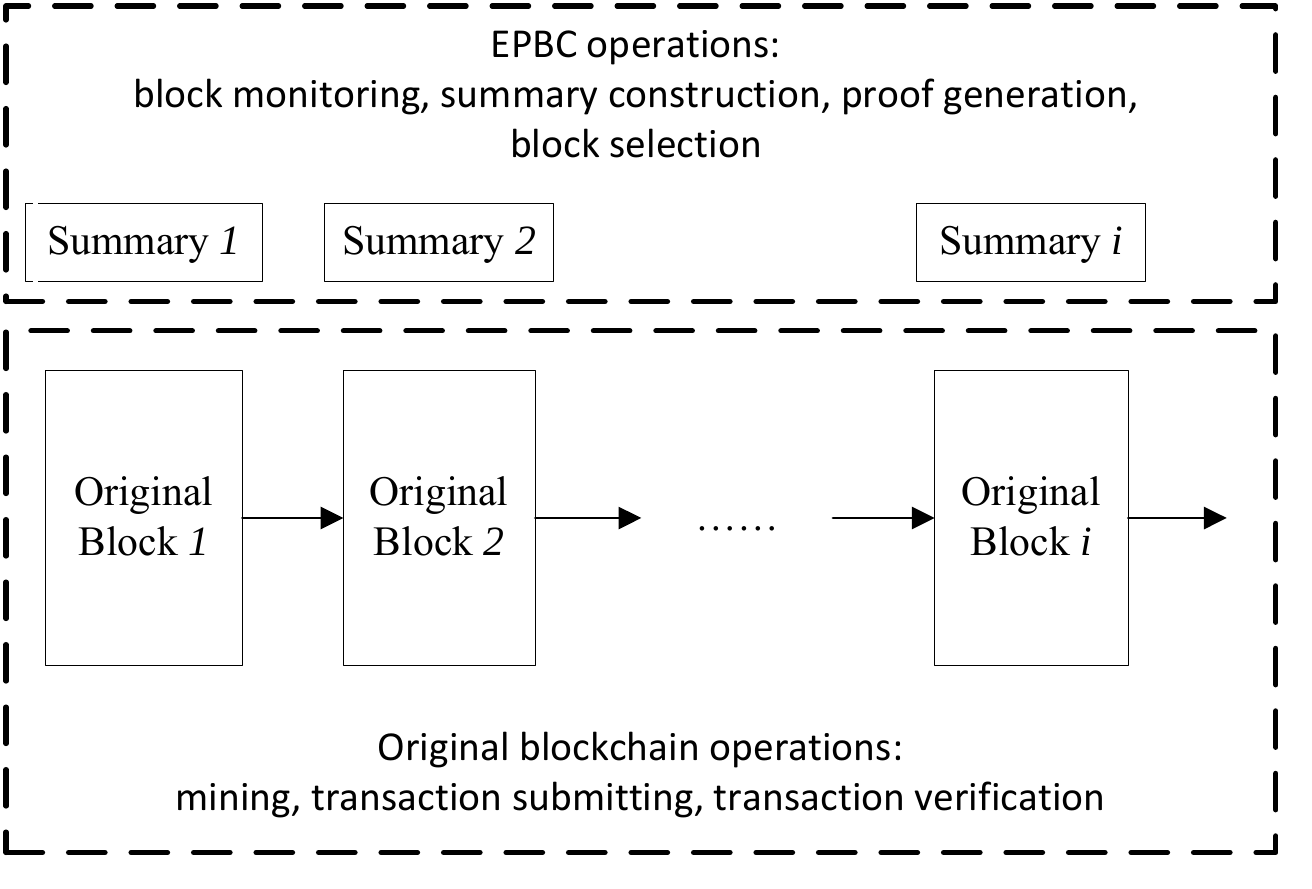}
  \caption{Overview of the integration of EPBC with an existing blockchain systems. A dedicated client is used to support EPBC related operations.}
  \label{fig-integration}
\end{figure}

Because a lot of public blockchain applications have been developed, it is useful to enable EPBC for these systems without modifying existing data structures and client. To achieve this goal, EPBC can work as a separate service layer on top of existing blockchain systems. \figurename~\ref{fig-integration} demonstrates the relationship between the existing blockchain system and the newly added EPBC service.

Specifically, a separate EPBC client with embedded parameters can be distributed to users who maintain the blockchain and play the role of a prover. Here parameters are values that used for blockchain summary construction. Summaries of the blockchain are not involved in mining, and users can use existing client to produce new blocks and achieve consensus on the blockchain. After the user decides to accept a new block, the EPBC client produces a new summary based on previous summary value and the new block, and stores the new summary locally. Note that summaries are determined by the blockchain itself so EPBC client does not need to run any consensus mechanism. If the user wants to reduce the time complexity of generating a proof, EPBC client can maintain the tree structure described in Section~\ref{subsec-tree-structure}.

\section{Experiments and Evaluation}\label{sec-imp-exp}
In this section, we describe the implementation and provide preliminary experimental results of EPBC. We focus on the block verification protocol because it is the core of EPBC.

\paragraph{Implementation and parameters.}
We implemented a prototype of the block verification protocol based on the MIRACL crypto library~\cite{miracl}. Since security of the protocol depends on the Strong RSA assumption,
we chose a 1,024 bits $N$ in the implementation. SHA256 was used for $\texttt{hash}()$.
We also set the height of $\mathcal{T}$ as 32. When a leaf is empty, its value is set to 1 and there is no need to store it.

\begin{figure}[!htbp]
  \centering
  \includegraphics[width=3.5in]{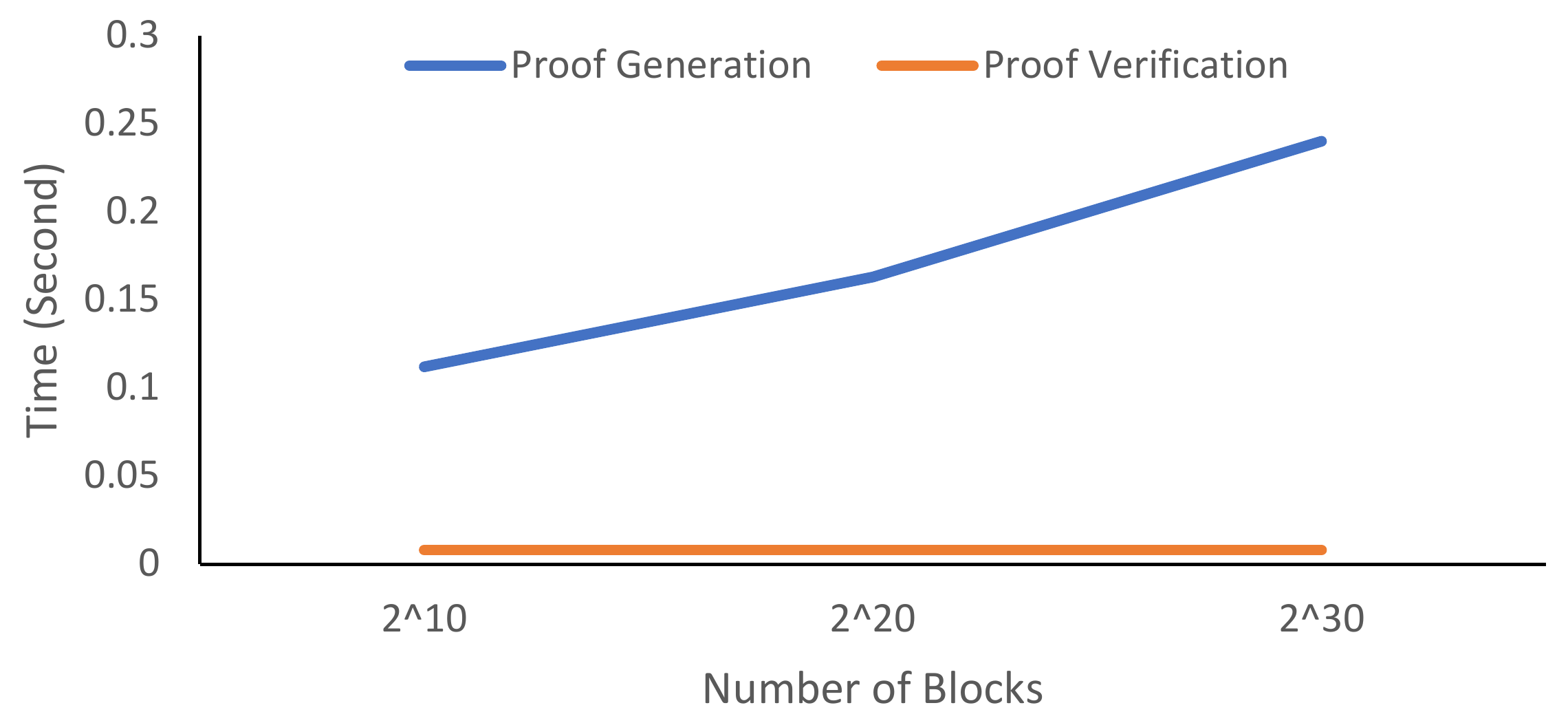}
  \caption{Preliminary experimental results of the block verification protocol using a low end Celeron CPU.}
  \label{fig-perfm}
\end{figure}

\paragraph{Experimental results.}
We conducted the experiments on a desktop with a low-end Intel Celeron 1017U processor, which has a similar Geekbench 4 score of Snapdragon 805 processor~\cite{geekbenchmark4}. The experimental results are summarized in \figurename~\ref{fig-perfm}, which shows that although the cost of proof generation depends on the size of the blockchain, the cost of proof verification is independent of the blockchain size.

As discussed in Section~\ref{sec-convert-operations}, some high-level operations like balance checking require the lightweight client to verify more than one blocks. This is not a problem in practice for the user using lightweight client because it only takes about 0.02 second to verify one block.

\section{Related Works}\label{sec-related}
EPBC only provides the mechanism for checking the validity of a given block and the transactions contained in the block. It does not consider how to determine which block(s) should be checked. It is proposed in BIP 37 to use a bloom filter to select potentially related blocks for verification~\cite{hearn2012connection}. The Bitcoin community proposes the UTXO (unspent transaction outputs) technology, which requires the user to store unspent transaction output information instead of transaction information. This reduces the storage cost but does not change the order of storage complexity~\cite{bishop2015review}.

Cryptographic accumulator was first developed by Benaloh and De Mare to achieve decentralized digital signature~\cite{benaloh1993one}. Bari{\'c} and Pfitzmann developed a collision-free accumulator and used it for fail-stop signatures without using any tree structure~\cite{baric1997collision}. Cryptographic accumulators are useful (e.g., constructing group signatures \cite{tsudik2003accumulating}). Dynamic cryptographic accumulator can further support adding/removing members~\cite{goodrich2002efficient}. These schemes do not consider features of blockchains, namely that every user has the privilege to construct blocks and generate proofs and lightweight users have very limited computational capability. Recently, e-cash systems such as ZeroCoin also utilizes cryptographic accumulators, but for a different purpose of information hiding~\cite{miers2013zerocoin}.

Another line of related research is storage verification in the cloud environment, and several related concepts were proposed, e.g., provable data possession~\cite{AtenieseBurnsCurtmolaEtAl2007} and proof of retrievability~\cite{zheng2011fair}. These schemes cannot be applied in our scenario because the lightweight users do not know the blockchain in advance and the blockchain keeps growing as new blocks are created and appended to it.

Both EPBC and SPV assume the records that are embedded into blocks are correct if the corresponding blocks are valid.
Some techniques that are applicable to SPV, such as bloom filter~\cite{gervais2014privacy}, are also applicable to EPBC.
Nevertheless, EPBC incurs only a constant amount of storage for the lightweight client, assuming the client cares about most recent transactions. This is significant because storing several block headers might be cheaper than storing the summary value.

\section{Conclusion}\label{sec-conclusion}
We have presented EPBC, a scheme for lightweight users to use blockchain-based applications without storing the entire blockchain while still able to verify the validity of blocks and transaction.
The basic idea is to ``compress'' a blockchain to a constant-size summary, which is the only data item a lightweight client needs to keep.
We analyzed the security of EPBC and preliminary experiments showed that it is practical.
EPBC can be adopted for blockchain-based applications, such as e-cash and smart contract systems.

\input{ref.tex}

\end{document}

%% file: ref.tex
\balance